\documentclass[aps, prl, 
twocolumn, superscriptaddress, showpacs, preprintnumbers]{revtex4}
\usepackage{amsmath,amssymb}
\usepackage{bm}

\newcommand{\av}[1]{\left\langle #1 \right\rangle} 
\renewcommand{\bar}[1]{\overline{#1}}%
\newcommand{\lam}{\lambda}

\newcommand{\vev}{\av{\phi}/V}
\newcommand{\nconf}{ {N_\text{conf}} }
\usepackage[dvipdfmx]{color}
\RequirePackage{ifpdf}
\ifpdf
  \usepackage[pdftex]{graphicx}
\else
  \usepackage[dvipdfmx]{graphicx}
\fi
\usepackage{cases}
\usepackage[utf8]{inputenc} 
\usepackage[english]{babel}
\begin{document}
%
%
\title{
Towards reduction of autocorrelation in HMC by machine learning
}
\author{Akinori Tanaka}
\email[]{akinori.tanaka@riken.jp}
\affiliation{Mathematical Science Team, RIKEN Center for Advanced Intelligence Project (AIP),1-4-1 Nihonbashi, Chuo-ku, Tokyo 103-0027, Japan}
\affiliation{Department of Mathematics, Faculty of Science and Technology, Keio University, 3-14-1 Hiyoshi, Kouhoku-ku, Yokohama 223-8522, Japan}
\affiliation{interdisciplinary Theoretical \& Mathematical Sciences Program (iTHEMS) RIKEN 2-1, Hirosawa, Wako, Saitama 351-0198, Japan}

\author{Akio Tomiya}
\email[]{akio.tomiya@mail.ccnu.edu.cn}
\affiliation{\small Key Laboratory of Quark \& Lepton Physics (MOE) and Institute of Particle Physics, Central China Normal University, Wuhan 430079, China}
\begin{abstract}
In this paper we propose new algorithm to reduce autocorrelation in Markov chain Monte-Carlo algorithms  for euclidean field theories on the lattice.
Our proposing algorithm is the Hybrid Monte-Carlo algorithm (HMC) with restricted Boltzmann machine.
We examine the validity of the algorithm by employing the phi-fourth theory in three dimension.
We observe reduction of the autocorrelation both in symmetric and broken phase as well.
Our proposing algorithm provides consistent
central values of expectation values of the action density and one-point Green's function with ones from the original HMC
in both the symmetric phase and broken phase within the statistical error.
On the other hand, two-point Green's functions have slight difference between one calculated by the HMC and one by our proposing algorithm
in the symmetric phase.
Furthermore, near the criticality, the distribution of the one-point Green's function differs from the one from HMC.
We discuss the origin of discrepancies and its improvement.
\end{abstract}

\maketitle


\section{Introduction}\label{section:introduction}
The dynamics of gluons and quarks is described by Quantum Chromo-Dynamics (QCD) but it has not been solved
analytically.
Since QCD is regularized by introducing an ultraviolet cutoff to
the spacetime \cite{Wilson:1974sk, Luscher:1976ms,Osterwalder:1977pc} (lattice QCD),
thus, we can evaluate physical observables 
using one of Markov Chain Monte-Carlo (MCMC) called the Hybrid Monte-Carlo algorithm (HMC) \cite{DUANE1987216}. 
Lattice QCD with HMC has been succeeded to reproduce important features of QCD:
the chiral symmetry breaking \cite{Fukaya:2007yv},
nuclear potentials \cite{Aoki:2009ji}
and QCD phase structure (\cite{Philipsen:2012nu,Ding:2015ona} and references therein).

A sequence of generated configurations by HMC are suffered from autocorrelation.
Long correlations between generated configurations reduce the effective number of configurations and it causes inefficiency of HMC.
As a related topic, the critical slowing down is becoming a problem in current lattice QCD calculations with HMC
 \cite{Schaefer:2009xx, Schaefer:2010hu}
because it actually induces long autocorrelation.
The autocorrelation in HMC is a vexing problem since HMC is based on the local update.
The long autocorrelation problem reminds us the Berlin wall problem in lattice QCD in the last decade \cite{Jansen:2003nt}.
It was solved by introducing multi-time step integrator \cite{Sexton:1992nu} and
the Hasenbusch mass preconditioning \cite{Hasenbusch:2001ne} to HMC.
For the current issue, we believe it is solved by introducing new idea also \cite{Hasenbusch:2017fsd,Cossu:2017eys,Bonati:2017woi}.

\begin{figure}[t]
\includegraphics[width =9cm]{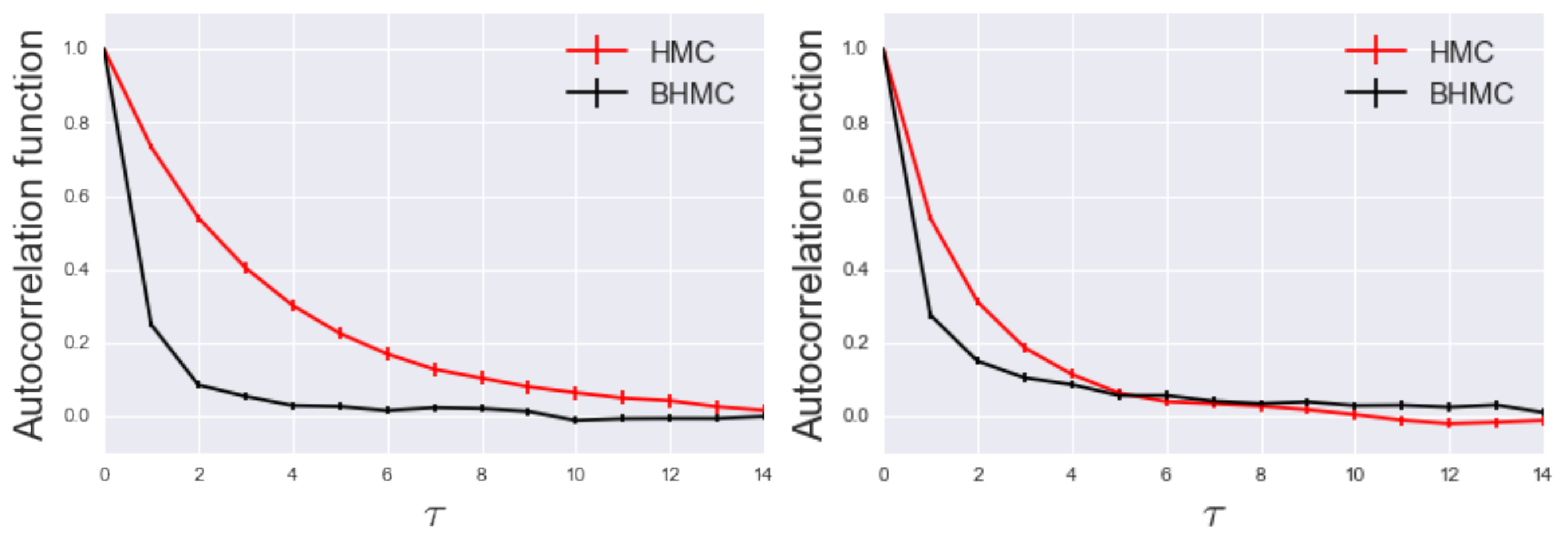}
\caption{
Left: Autocorrelation function for the condensate in the symmetric phase.
Right: The same plot but in the broken phase.
Black lines represent autocorrelations for our Boltzmann machine supported HMC.
}
\label{fig:autocorrelations}
\end{figure}

Recently, striking idea, Self Learning Monte-Carlo (SLMC) algorithms, are suggested \cite{liu2017self, liu2016self, xu2016self, nagai2017self}
for several models of condensed matter physics.
They assume effective Hamiltonian with some couplings as free parameters and determine
the couplings through linear regression with the original Hamiltonian.
After the regression, they use the trained effective Hamiltonian to generate configurations
and they obtain a sequence of configurations with shorter autocorrelation.
Similar but more radical approach can be found in \cite{huang2017accelerated} which replaces the effective Hamiltonian with the Binary Restricted Boltzmann machine (BRBM) which is one of the well known architectures called \textit{generative models} in the machine learning community. 
Compared to the previous effective Hamiltonian approach, using generative models might be better in a following sense.
In SLMC, first we need to prepare the most generic form of the effective Hamiltonian, and after some experiments, we can determine the important couplings to fit the actual Hamiltonian.
On the other hand, we do not need to perform such experiments in generative models.
It is general-purpose architecture to fit arbitrary probability density (Ch. 20 in \cite{Goodfellow-et-al-2016}), and in fact, it is reported that BRBM can sample the Ising configurations approximately \cite{morningstar2017deep} even near the criticality.
So, replacing effective Hamiltonian in SLMC to certain generative model may be a good idea to solve critical slowing down in lattice QCD.

Towards this goal, we modify HMC by introducing real-valued RBM called 
Gaussian-Bernoulli Restricted Boltzmann Machines (GRBM).
We call it {\it BHMC} (Boltzmann machine assisted HMC) in short.
To examine our proposal's validity, we employ interacting real scalar field theory in three dimensional discretized spacetime.
The theory is described by the action, 
\begin{align} 
S [\phi] = \sum_n^N \left[ -\frac{1}{2} \phi_n \Delta \phi_n + \frac{m^2}{2}\phi^2_n+\frac{\lambda}{4!}\phi^4_n  \right].
\label{action}
\end{align}
Our notation and setup
are summarized in appendix A.

\newcommand{\data}{$10^4$}
\begin{table}[t]
\begin{tabular}{c||c|c|c|c|c} 
Alg.&
Phase& $\nconf $ & $\vev $ & $\av{S}/V$ & $\tau_\text{int}$ \\\hline\hline
HMC&Symmetric& \data &  
0.00$\pm$0.05 & 
0.48$\pm$0.03 &
4.4$\pm$0.3\\ 
&Broken& \data &  
-3.94$\pm$0.04& 
-2.70$\pm$0.03& 
2.8$\pm$0.2\\ 

\hline
BHMC &
Symmetric& \data & 
0.00$\pm$0.04 & 
0.50$\pm$0.04 &
2.0$\pm$0.1\\ 
&Broken& \data & 
-3.95$\pm$0.03 & 
-2.73$\pm$0.04& 
2.5$\pm$0.2 
\end{tabular}
\caption{
{\it BHMC} in the Alg. column is our proposing algorithm (HMC+GRBM). $V$ is the system volume $8^3$ and
$\nconf$ is the number of configurations.
The integrated autocorrelation time
$\tau_\text{int}$ is defined by the expectation value for the spacetime averaged one-point Green's function.
Here the initial configuration both for HMC and BHMC is prepared from a configuration which is well thermalized configuration from HMC.
\label{tab:data}}
\end{table}
In order to confirm validity of BHMC, we compare following quantities calculated by configurations generated by both of the original HMC and BHMC.
First is 
the expectation value of the action density for \eqref{action}, $\av{ S }/V$,
\begin{align}
\bar{ S }/V = \frac{1}{V} \frac{1}{\nconf } \sum_{c=1}^{\nconf } S [\phi^{(c)}],
\label{s_result}
\end{align}
where $V=N_xN_yN_t$ is the spacetime volume, $\nconf $ is the number of configurations and $\phi^{(c)}$ is $c$-th configuration.
Second is 
the vacuum expectation value of one-point Green's function (vev) $\vev$,
\begin{align}
\bar{ \phi }/V  = \frac{1}{\nconf} \sum_{c=1}^{\nconf} \frac{1}{V} \sum_{n}^N \phi_{n} ^{(c)},
\label{vev_result}
\end{align}
where $\phi_{n} ^{(c)}$ is a field value at point $n$ for $c$-th configuration.
Third quantity is 
the two-point Green's function with zero momentum projection,
\begin{align}
\bar{G}(t) =
\sum_{dx, dy = 1}^{N_x, N_y}
\frac{1}{\nconf }
\sum_{c = 1}^{\nconf }
\frac{1}{V}
\sum_{n} ^N
\phi_{t+n_t, dx+n_x, dy+n_y}^{(c)}
\phi_{n}^{(c)},
\label{gt_result}
\end{align}
In this paper, we call ${G}(t)$ the two-point function.
In the broken phase for this model, there are no Nambu-Goldstone modes,
thus we employ the connected part of the two-point function,
\begin{align}
\bar{G}_\text{con}(t) = \bar{G}(t) - N_x N_y | \bar{ \phi }/V |^2,
\end{align}
in stead of $\bar{G}(t)$ itself and examine ${G}_\text{con}(t)=0$.
These quantities are the important to examine
 the proposed algorithm whether it provides ``physically legal'' configurations or not.

Besides these quantities, we calculate the approximated normalized autocorrelation function $\bar\rho(\tau)$
\cite{Wolff:2003sm, Madras:1988ei}
(See appendix B).
Particularly we focus on the autocorrelation of vev. 
In addition, we calculate the integrated autocorrelation $\tau_\text{int}$ from $\bar\rho(\tau)$.

\vspace{.5cm}
This paper is organized as follows.
In next section, we introduce our HMC based algorithm which is assisted by GRBM. 
After that, we compare numerical results calculated by the original HMC and our algorithm in two phases.
Table \ref{tab:data} shows a part of results, and looks consistent.
But for skeptical readers, we go further and show discrepancies with HMC also.
%
In the final section, we summarize and discuss these discrepancies and its improvement. 
In addition, we address on a issue of our algorithm near the criticality.

\section{Boltzmann machine assisted Hybrid Monte Carlo (BHMC) algorithm}
\begin{figure}
\includegraphics[width =6.5cm]{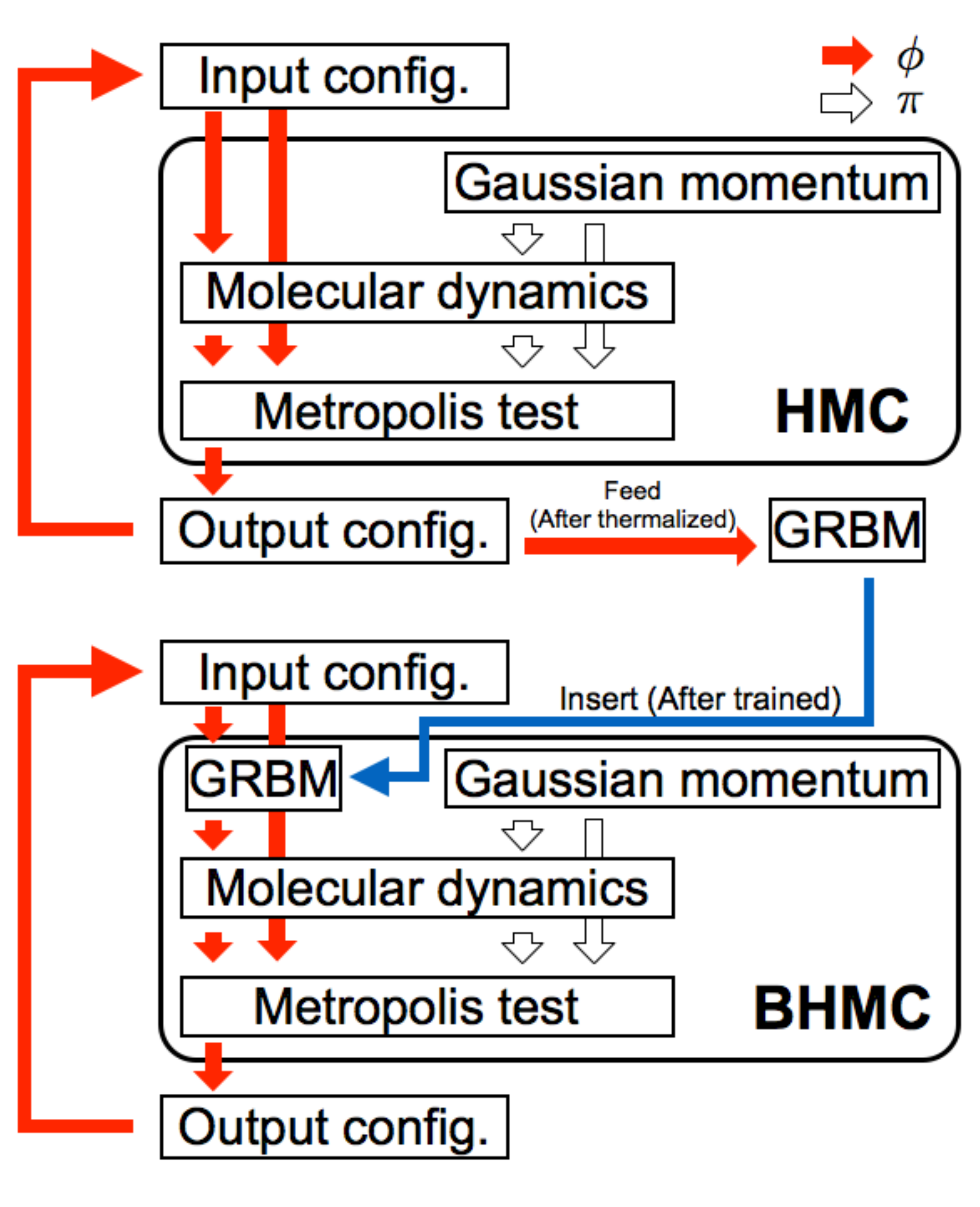}
\caption{
Our proposal for update (BHMC).
Gaussian-Bernoulli RBM (GRBM) is added to the original HMC after training.
}
\label{fig:idea}
\end{figure}

In order to introduce our proposal, let us briefly review procedures of the original HMC \cite{DUANE1987216}
(see appendix B for details).
HMC consists of three steps:
\begin{enumerate}
\item {\it Momentum refreshment}: Generate a set of $\pi_n$ for every points $n$ from the Gaussian distribution. \label{item:appendix_HMC_HB}
\item {\it The molecular dynamics}: Fields $(\phi, \pi)$ are evolved to $(\phi', \pi')$ using the canonical equations of motion for $H_\text{HMC}[\phi,\pi] = S[\phi] + \frac{1}{2} \pi^2$. 
\label{item:appendix_HMC_MD}
\item {\it The Metropolis test}: Chooses a next pair from candidates $(\phi', \pi')$ and $(\phi, \pi)$ using $H_\text{HMC}[\phi,\pi]$. \label{item:appendix_HMC_Met}
\end{enumerate}
This algorithm is schematically shown in the top half of Figure \ref{fig:idea}.
In step \ref{item:appendix_HMC_MD},
a candidate configuration is generated by solving equations of motion for $H_\text{HMC}[\phi,\pi]$.
Step \ref{item:appendix_HMC_Met} ensures generated configurations obey the distribution with the Boltzmann weight $e^{-H_\text{HMC}[\phi,\pi]}$.

In total, we can generate a sequence of configurations $\phi$ which obeys the 
distribution $e^{-S[\phi]}$ by repeating above three steps.
However updates of configurations are essentially done by evolution according to the equation of  motion for $H_\text{HMC}$ only
(step \ref{item:appendix_HMC_MD}). Namely, local updates for the field but this seems unavoidable from the gauge symmetry and psudofemion
formalism in the case of lattice QCD and this leads long autocorrelation.
Toward solving this problem, we introduce a Gaussian-Bernoulli Restricted Boltzmann Machine
(GRBM) into the algorithm (Bottom half of Figure. \ref{fig:idea}).

\vspace{.5cm}
GRBM can be regarded as a physical system composed by dynamical field $\phi$ and auxiliary \textit{binary} field $h$ with the following Hamiltonian:
\begin{align}
H^\text{GRBM}
_\theta[\phi, h]
=
\sum_{n}
\frac{(\phi_n - a_n)^2}{2 \sigma_n^2}
-
\sum_{j}
b_j h_j
-
\sum_{n, j}
\frac{\phi_n}{\sigma_n} w_{nj} h_j,
\label{RBM_H}
\end{align}
where $\theta = (a_n, \sigma_n, b_j, w_{nj})$.
In contrast to usual statistical physics problem which discuss property of the system with fixed $\theta$,
we will \textit{determine} appropriate $\theta$ by the following steps.
\begin{itemize}
\item Preparing a teacher data $\mathcal{D} \sim e^{-S[\phi]}/\mathcal{Z}_\text{lat}$ which we would like to mimic (top half of Figure. \ref{fig:idea}).
\item Updating parameters $\theta$ in \eqref{RBM_H} via \textit{contrastive divergence} method \cite{hinton2006training, bengio2009justifying} (See appendix C for a brief review.).
\end{itemize}
Throughout this paper, we prepare teacher data $\mathcal{D}$ by the original HMC.
In this sense our algorithm is not completely independent to the original HMC, but once we finish the training, we can use GRBM as a sampler of $\phi$ via block Gibbs sampling \eqref{BG}.

Thanks to the intermediate hidden states in block Gibbs sampling, \textit{i.e.} states of $h$, the integrated autocorrelation time for the block Gibbs sampling is expected to be short.
However, it sounds dangerous just relying on block Gibbs sampling, 
so we apply the molecular dynamics and the Metropolis test for sampled configurations.
In summary our proposal for improvement is replacing 2. in HMC algorithm to
\begin{enumerate}
\setcounter{enumi}{1}
\item Generating new $\phi_\text{GRBM}$ via block Gibbs sampling by $H_\theta^\text{GRBM}$ from the given $\phi$, and giving pair $(\phi', \pi')$ from the molecular dynamics development by $H_\text{HMC}$ with $(\phi_\text{GRBM}, \pi)$. 
\end{enumerate}
This is schematically shown in the bottom Figure \ref{fig:idea}.

\section{Experiment and Result}
\begin{figure}[t]
\includegraphics[width =9cm]{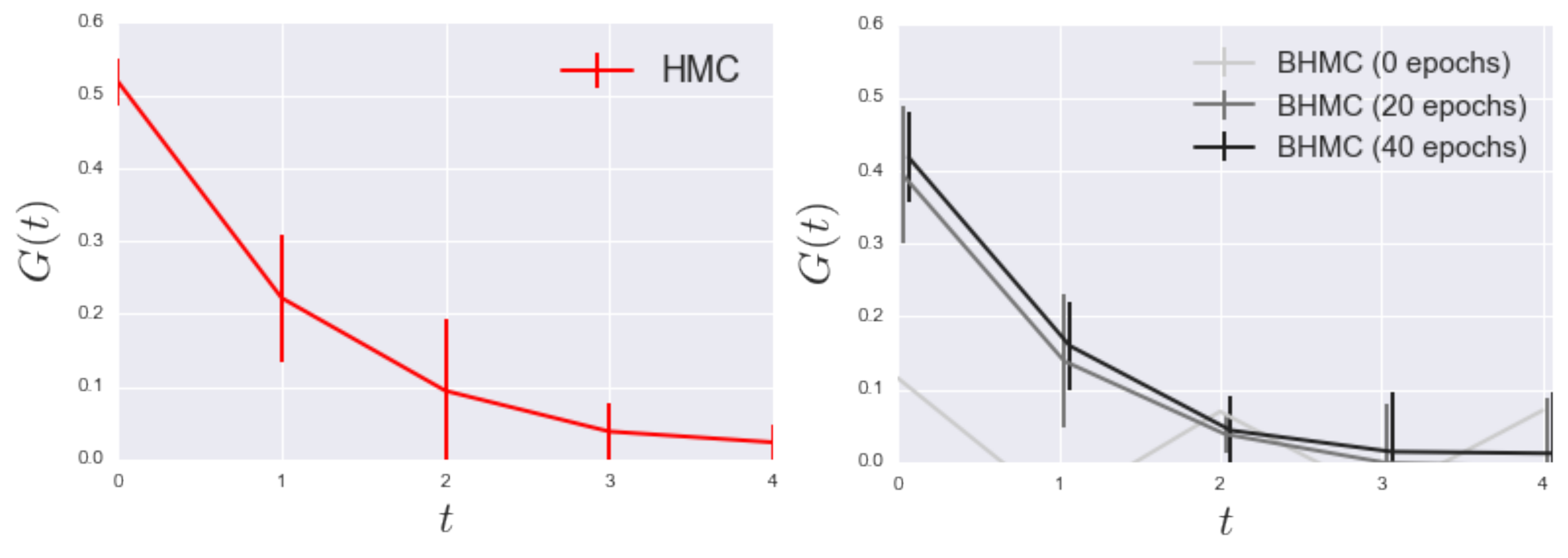}
\caption{
Two point function ${G}(t)$ for generated samples in the symmetric phase.
Left: HMC,
Right: BHMC
}
\label{fig:2pts}

\vspace{.5cm}
\includegraphics[width =9cm]{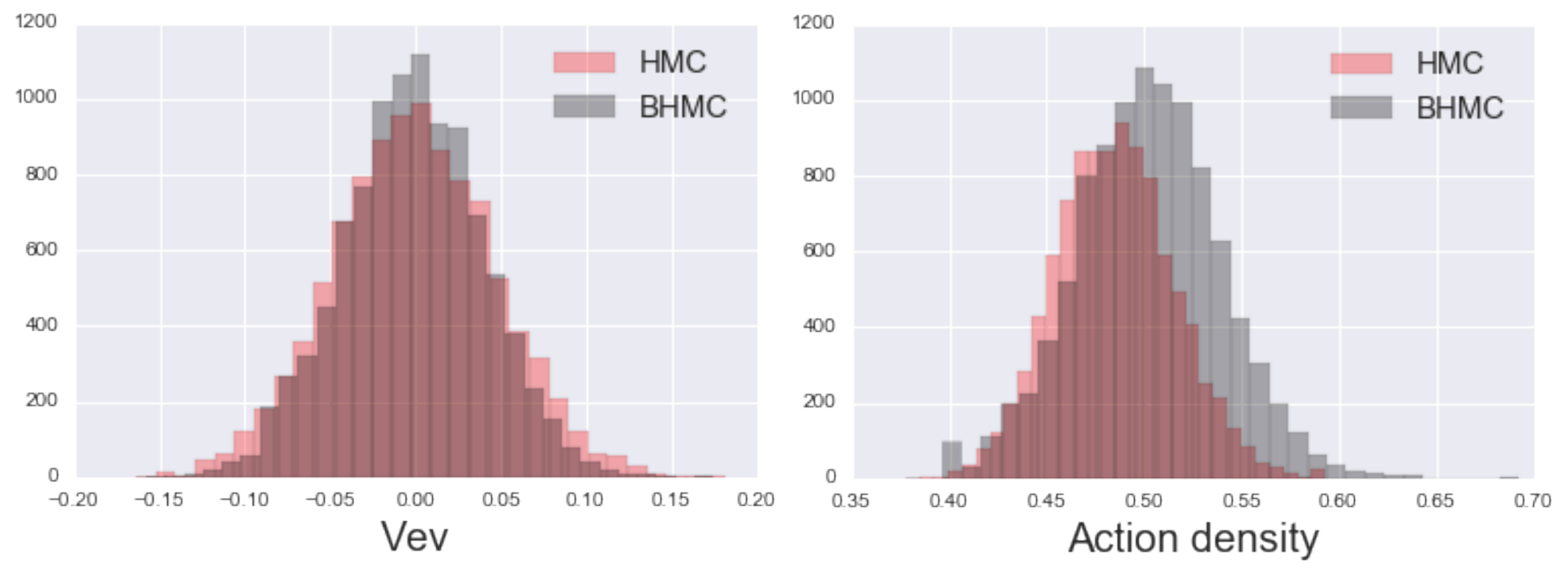}
\caption{
Histograms for generated samples in the symmetric phase.
Left: vev for HMC(red) and trained BHMC (black),
Right: the action density for HMC(red) and trained BHMC (black).
}
\label{fig:vasym}
\end{figure}

\newcommand{\numData}{10^{4}}
\newcommand{\numBatch}{10^{2}}
\newcommand{\epoch}{40}
\newcommand{\rateCD}{10^{-3}}
\newcommand{\rateWD}{10^{-4}}
\newcommand{\discardNum}{10^{3}}

\paragraph{Preparation of teacher data $\mathcal{D}$:}
We prepare $\nconf = \numData$ teacher configurations for training on
$ N = (8, 8, 8)$ lattice by running HMC with cold start.
We choose $\Delta\tau = 0.2$ for the integration step size in the molecular dynamics.
Here earlier $\discardNum$ configurations are discarded and after that
we start storing the teacher configurations.

\paragraph{Training details:}
The learning rate  in \eqref{learning} are taken as $\epsilon = \rateCD, \eta = \rateWD$.
$a_n$ and $b_j$ are initialized as zero vectors, and $\sigma_n$ is set to one.
$w_{nj}$ is sampled from Gaussian distribution $\mathcal{N} (0 | 0.01)$ as recommended in \cite{hinton2010practical}. 
We train our model by \textit{minibatch learning} in following way.
\begin{enumerate}
\renewcommand{\labelenumi}{(\roman{enumi})}
\item{
We randomly divide teacher data as direct sum $\mathcal{D} = \cup_b \mathcal{D}_\text{mini}^{(b)}$ where each $\mathcal{D}_\text{mini}^{(b)}$ includes $\numBatch$ samples.}
\item{
We repeat updating process for $b$.
Each update is performed by using mean values on $\mathcal{D}_\text{mini}^{(b)}$ of $\delta \theta$. 
}
\end{enumerate}
We call the pair of procedure (i) and (ii) as \textit{epoch}.
We train GRBM with $\epoch$ epochs.
\paragraph{Test details:}
After training, we run BHMC to generate $10^4$ configurations.
In addition, we perform the original HMC to generate $10^4$ configurations for a reference.
We use $\Delta \tau = 0.2$ in both molecular dynamics.
Both of algorithms start with the same initial thermalized configuration.
By using each configuration, we compute vev \eqref{vev_result}, the action density \eqref{s_result}, two-point function \eqref{gt_result}, and examine BHMC can generate sane configurations or not.
In addition we calculate autocorrelation function \eqref{autocorrelation_function}, integrated autocorrelation \eqref{tau_int}, and examine BHMC can provide small autocorrelation time or not.

We perform experiments in two phases of the $\phi^4$-theory, namely
the symmetric phase and broken phase as follows.

\subsubsection{The symmetric phase}

\newcommand{\stepTrain}{10,000}
\newcommand{\stepTest}{10,000}
\newcommand{\accTrain}{70\%}
\newcommand{\accVal}{70\%}
\newcommand{\accTest}{70\%}
\newcommand{\msqSym}{0.8}
\newcommand{\lamSym}{0}

We take $m^2 =\msqSym$, $\lam = \lamSym$.
Acceptance rate for each MCMC is, 
\begin{align}
&\text{BHMC(0 epoch): } 0.0 \%, \notag \\
&\text{BHMC(20 epoch): } 38.97 \%, \notag \\
&\text{BHMC(40 epoch): } 84.49 \%, \notag \\
&\text{HMC : }71.66 \% \notag
.
\end{align}
As easily noticed, values of acceptance rate for BHMC increases with increasing epochs, {\it i.e.} the number of training iterations.
This is a signal of success on training GRBM.

The integrated autocorrelation times $\tau_\text{int}$, 
vev and the action density are summarized in Table \ref{tab:data}.
They are all fit in the ones in legal configuration, and BHMC provides configurations having shorter autocorrelation as expected.
This can be seen from the behavior of  the autocorrelation functions (Left panel in Figure \ref{fig:autocorrelations}).

We plot the two-point function ${G}(t)$ for the original HMC and BHMC in Figure \ref{fig:2pts}
and \eqref{eq:G(t),symmetric}.
\begin{align}
\left. \begin{array}{c|c|c|c|c|c|}
 & G(0) & G(1) & G(2) & G(3) & G(4) \\
\hline
\text{HMC} & 
0.52(3)&
0.22(8)&
0.09(9)&
0.04(4)&
0.02(3)
\\
\text{BHMC} & 
0.42(6)&
0.16(6)&
0.04(5)&
0.01(8)&
0.01(8)
\\
\end{array} \right.
\label{eq:G(t),symmetric}
\end{align}
In the right panel of Figure \ref{fig:2pts}, the darker line corresponds calculated based on more trained GRBM.
The magnitude of ${G}(t)$ for BHMC is slightly smaller than the one for the HMC for all $t$.

In addition, we plot histograms of vev and the action density in Figure \ref{fig:vasym}.
The histogram of vev seems sanity.
On the other hand, the distribution of the action density of BHMC have slight discrepancy between the ones of HMC
conservatively.

\begin{figure}[t]
\includegraphics[width =9cm]{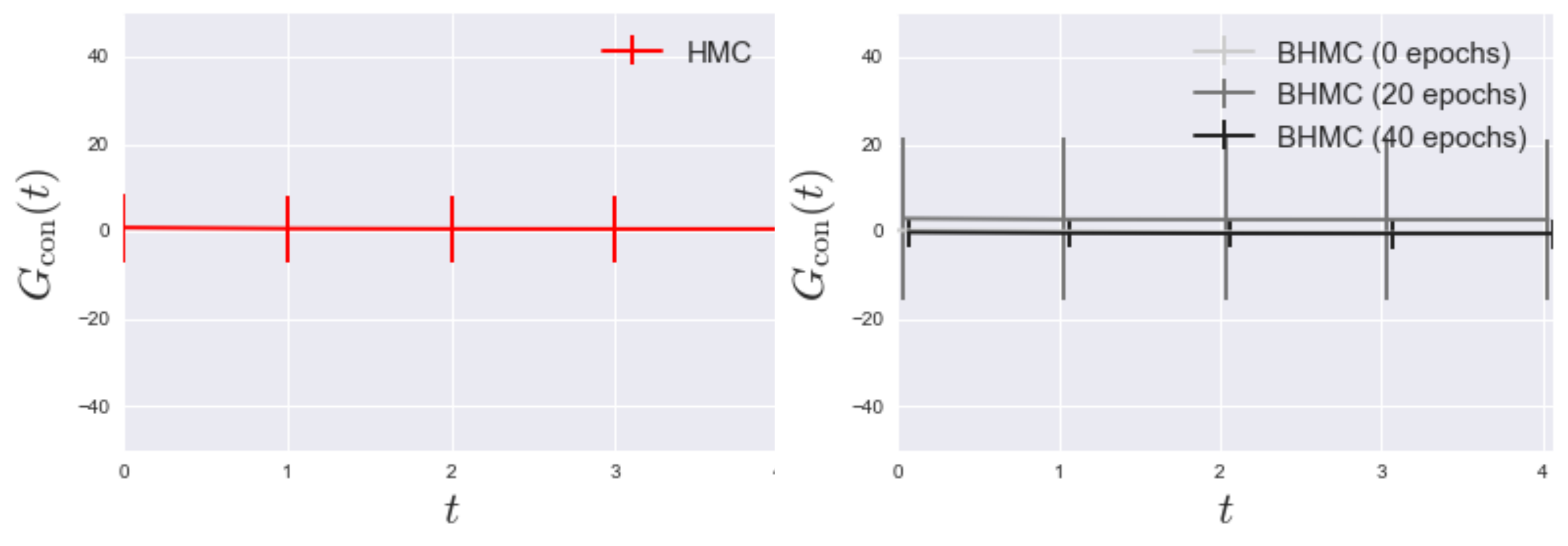}
\caption{
Two point function $G_\text{con}(t)$ for generated samples in the broken phase.
Left: HMC, Right: BHMC.
}
\label{fig:2ptb}

\vspace{.5cm}
\includegraphics[width =9cm]{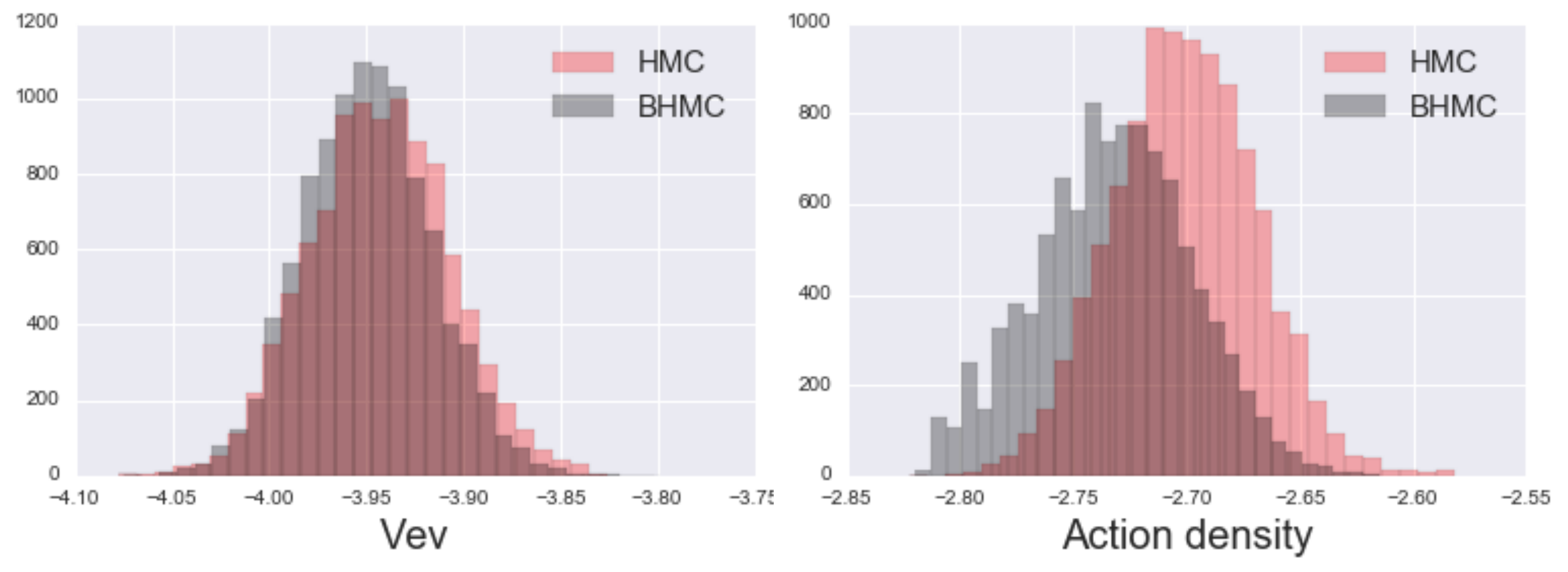}
\caption{
Histograms for generated samples in the broken phase.
Left: vev for HMC (red) and trained BHMC (black),
Right: the action density for HMC (red) and trained BHMC (black).
}
\label{fig:vabro}
\end{figure}
\subsubsection{The broken phase}

\renewcommand{\stepTrain}{10,000}
\renewcommand{\stepTest}{10,000}
\renewcommand{\accTrain}{70\%}
\renewcommand{\accVal}{70\%}
\renewcommand{\accTest}{70\%}
\renewcommand{\msqSym}{-0.8}
\renewcommand{\lamSym}{0.3}

We take $m^2 =\msqSym$, $\lam = \lamSym$.
Acceptance rate for each MCMC is, 
\begin{align}
&\text{BHMC(0 epoch): } 0.0 \%, \notag \\
&\text{BHMC(20 epoch): } 13.36 \%, \notag \\
&\text{BHMC(40 epoch): } 69.37 \%, \notag \\
&\text{HMC : }71.19 \% \notag
.
\end{align}
Comparing to BHMC in the symmetric phase, the acceptance rate does not large. 

The integrated autocorrelation time $\tau_\text{int}$, 
vev and the action density are summarized in Table \ref{tab:data}.
$\tau_\text{int}$ for BHMC is also shorter than one for HMC as well as the symmetric phase.
This can be seen from the behavior of  the autocorrelation functions (Right panel in Figure \ref{fig:autocorrelations}).

The connected part of two-point function is showed in Figure \ref{fig:2ptb}.
The two-point function by BHMC is consistent with the one by the original HMC,
{\it i.e.} it is zero for all $t$, within the statistical error. 

The histogram for vev is showed in the left panel in Figure \ref{fig:vabro},
which looks consistent with one by the HMC.
On the other hand,
the histogram for the action density is showed in the right panel in Figure \ref{fig:vabro},
which has discrepancy to one for the HMC.

\section{Summary and Discussion}
In this paper we have introduced an algorithm based on HMC with GRBM.
We have observed reduction of the autocorreation time both in the symmetric and broken phase.

\begin{figure}[t]
\includegraphics[width =5cm]{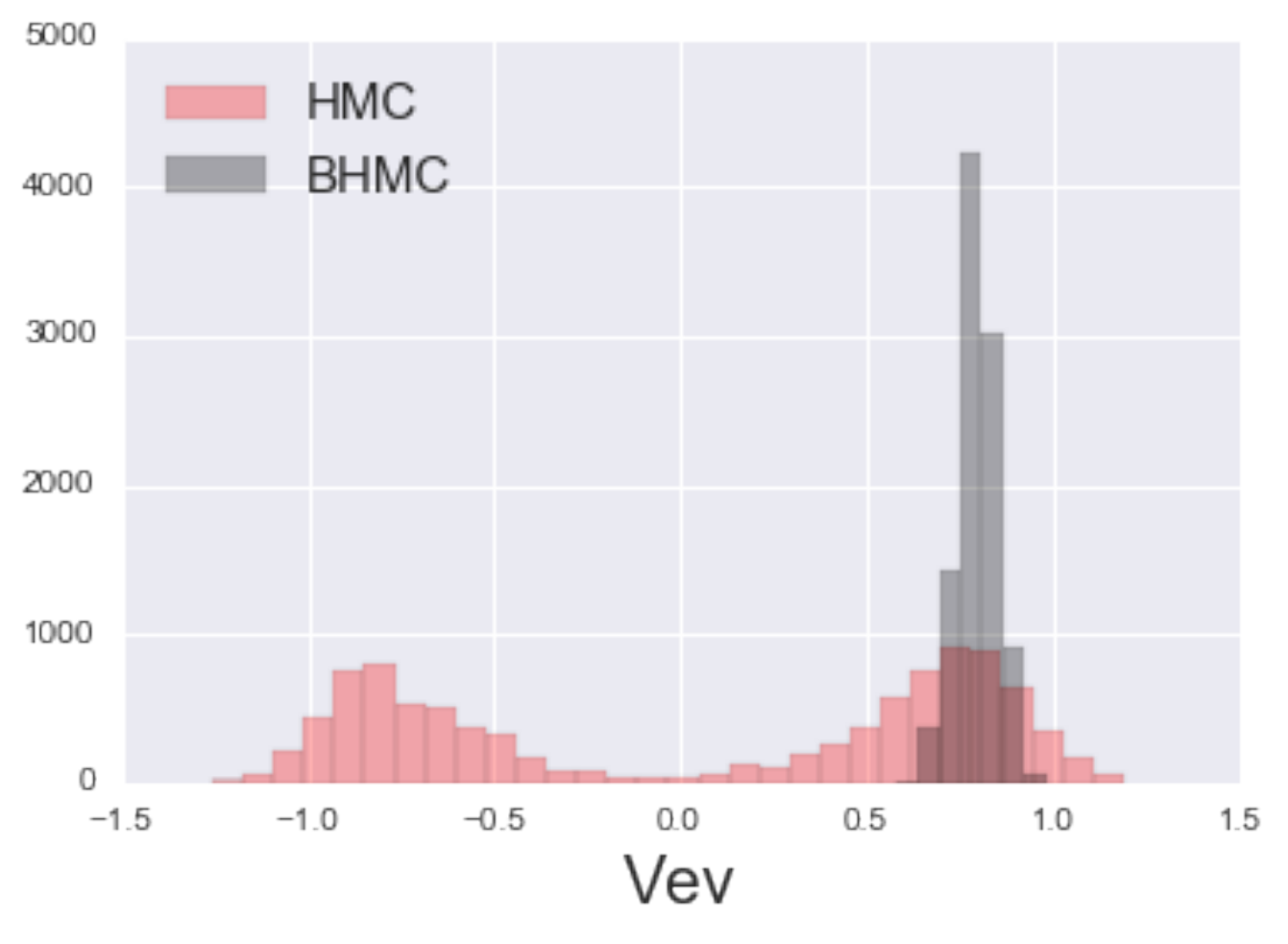}
\caption{
Histogram for vev near the criticality.
Red: HMC, Black: BHMC.
}
\label{fig:critical}
\end{figure}

Let us leave here some comments.
First, we can conclude that our new algorithm provides well approximated configurations because 
the action density and vev are consistent with ones by HMC (Table \ref{tab:data}) without fitting the Hamiltonian/action itself.
However, observables have different distributions.
In fact, similar phenomena are reported by \cite{morningstar2017deep} in Ising model.
This discrepancy has to be solved for practical use.
Second, to attack the critical slowing down problem in lattice QCD, 
we need to overcome the long autocorrelation problem \textit{near the criticality}, but our new algorithm is poor to sample from such phase boundary.
The reason is simple because the Hamiltonian \eqref{RBM_H} approximates the data by the Gaussian distributions for each spacetime point.
In symmetric/broken phase, the actual distribution for vev $\vev$ has single peak and it is reasonable to use \eqref{RBM_H} to approximate it.
On the other hand, near the criticality, it has double peaked distribution, which is not suitable for our formalism (Figure \ref{fig:critical}).

In order to represent rich distributions by this kind of frame work without breaking short autocorrelation,
one naive idea is adding more hidden layers to our GRBM.
In fact, it is known that the three-layered Boltzmann machines exceed RBM \cite{hinton2006fast}.
It may work, but we need another heavy MCMC to sample from deep Boltzmann machines.
Alternative idea is using neural networks.
Neural network itself is a deterministic architecture, and one may suspect its effectiveness to approximate probability density.
Recently, however, generative models based on neural networks \cite{goodfellow2014generative, arjovsky2017wasserstein} exceed conventional energy-based models like RBMs.
It uses simple fixed noise $z \sim p(z)$, and train the networks $G_\theta (z)$ regarding itself as sampling of target configuration $\phi$.
After the training of $G_\theta$, it is proved that the resulting distribution on $\phi = G_\theta (z)$ is identical to the true distribution.
Here MCMC is not needed, thus no autocorrelation problem at all.

If we want to use generative models as lattice QCD sampler, we must guarantee
the gauge symmetry of a probability distribution for the model.
This is because,
configurations which are generated by a algorithm must obey a conclusion of Elitzur's theorem \cite{Elitzur:1975im}, namely
expectation values of gauge variant quantities must be zero.
For example, Boltzmann machines do not have such mechanism.
In this sense, 
we may need to develop {\it generative models for gauge field configurations},
namely it can generate gauge field with appropriate redundancy.
Of course, if we fix the gauge, conventional generative models might work. 
However, from a viewpoint of efficiency, we believe that such models are needed.

\section*{Acknowledgement }
Akio Tomiya would like to thank to 
Taku Izubuchi, Yuki Nagai, and Testuya Onogi
for discussion in early stage of this work.
The work of Akinori Tanaka was supported by the RIKEN Center for AIP and the RIKEN iTHES Project.
Akio Tomiya was fully supported by Heng-Tong Ding.
The work of Akio Tomiya was supported in part by NSFC under grant no. 11535012.

\appendix

\section{Appendix A. $\phi^4$ theory: Physics and Notation \label{app:phi4} }
In this appendix, we explain physical aspects and Monte-Carlo calculation of $\phi^4$ theory to introduce our notation.
It is described by a discretized action \eqref{action},
\begin{align*} 
S[\phi] = \sum_n^N \left[ -\frac{1}{2} \phi_n \Delta \phi_n + \frac{m^2}{2}\phi^2_n+\frac{\lambda}{4!}\phi^4_n  \right],
\end{align*}
where $n = (n_t,n_x,n_y)$.
$N = (N_t,N_x,N_y)$ represents the size of the system and $n_\mu$ is an integer in a range $[0,N_\mu)$.
$\Delta$ is  Laplacian on the discrete spacetime without any improvement. 
$m^2$ and $\lambda$ are real parameters and $m^2$ can be negative.
In general, interacting scalar filed theories in three dimension are allowed to contain $\phi^6$ term in the action by the renormalizability but we just omit it for simplicity.
$\phi_ n$ is enjoined the periodic boundary condition for each direction. 

The expectation value of an operator $O$ is defined by the euclidean path integral,
\begin{align}
\av{O} = \frac{1}{\mathcal{Z}_\text{lat} } \int \mathcal{D} \phi \; O[\phi] \; e^{-S[\phi]  } \label{eq:lattice_QCD_original_path_int},
\end{align}
where $\mathcal{D}\phi = \prod_n^N d\phi_n$ and $\mathcal{Z}_\text{lat} = \int \mathcal{D} \phi \; e^{-S[\phi]  }$.
In the other words, realization probability for a certain configuration $\phi$ is given by,
\begin{align}
p_\text{lat}[\phi]= \frac{e^{-S[\phi]} }{\mathcal{Z}_\text{lat} } .
\end{align}
The expectation value is estimated by a Markov Chain Monte-Carlo algorithm (MCMC),
\begin{align}
\av{O}
&\approx 
\bar{O}
\equiv
\frac{1}{\nconf } \sum_{c=1}^{\nconf } O[ \phi^{(c)} ] ,
\end{align}
where $\phi^{(c)} = \{ \phi_n \}_c$ represents $c$-th configuration and $\nconf $ represents the number of configuration which are used in measurements of $O$.
The expectation value $\av{O} $ and its evaluation $ \bar{O}$ is related by,
\begin{align}
\av{O} = \bar{O} \pm \delta O,
\end{align}
where $\delta O$ is the statistical error of MCMC.
We estimate $\delta O$ by the standard deviation for vev and the action density,
and by the Jackknife method for two-point functions.

One of important observables is the vacuum expectation value of the field value $\phi$,
\begin{align}
\av{\phi}/V \equiv  \frac{1}{V} \sum_n^N
\frac{1}{\mathcal{Z}_\text{lat} } \int \mathcal{D} \phi \; \phi_n \; e^{-S[\phi]  },
\end{align}
where $V=N_t N_x N_y$ is the spacetime volume.
The action has $Z_2$ symmetry {\it i.e.} the action is invariant under a transformation $\phi_n \to -\phi_n$.
However the vacuum expectation value $\av{\phi}$ can acquire nonzero value which is controlled by parameters in the action $m^2$ and $\lambda$.
A phase is called the symmetric phase if $\av{\phi} = 0$,
and the other phase is called the broken phase.

The two-point function with zero momentum projection is define in following way,
in the continuum field theory.
The original definition of two-point function is,
\begin{align}
{G(r,r_0)}
&= \av{\phi(r)\phi(r_0)},
\end{align}
where $r=(t,x,y)$ and $t$ is the imaginary time.
By using the translational symmetry, it is written as a function of the distance,
\begin{align}
{G(r,r_0)} &= {G(r - r_0)}.
\end{align}
In order to enlarge the statistics, we take spacetime average, 
\begin{align}
{G(r)} = \frac{1}{V} \int d^3 r_0 \; {G(r - r_0)}.
\end{align}
Performing the Fourier transformation to $x,y$,
\begin{align}
{G(t,p)} = \int dx dy \; {G(r)} \; e^{i (x,y)\cdot \vec{p} }.
\end{align}
Finally, a limit $\vec{p}\to \vec{0}$ gives the two point function with zero momentum projection ${G(t)}={G(t,p=0)}$.
In general, the two-point function ${G(t)}$ is related to the lightest mass $m_0$ of propagating modes in the symmetric phase,
${G(t)} \sim e^{-m_0 t}$ for enough large $t$. 

Two-point Green's function can be decomposed into the connected
part and disconnected part,
\begin{align}
G(r) = {G_\text{con}(r)} - \left| \av{\phi(r)} \right|^2.
\end{align}
In terms of the zero momentum projected one,
\begin{align}
G_\text{con}(t) = G(t) - {N_x N_y} |\av{\phi}/V|^2.
\end{align}
For the broken phase, we use this subtracted one
and we expect $G_\text{con}(t)=0$ because no Nambu-Goldstone modes
in the broken phase for this model.

\section{Appendix B. HMC \label{app:HMC}}
Here we review conventional HMC \cite{DUANE1987216}.
Essentially, HMC is based on following expression which is equivalent to \eqref{eq:lattice_QCD_original_path_int},
\begin{align}
\av{O} = \int \mathcal{D} \phi \mathcal{D}\pi \; O[\phi] \; e^{-S[\phi]  - \frac{1}{2} \pi^2 } \Big/  \int \mathcal{D} \phi \mathcal{D}\pi \; e^{-S[\phi]  - \frac{1}{2} \pi^2 },
\end{align}
where $\pi^2 = \sum_n^N \pi_n^2$ is a real scalar auxiliary field, called (fictitious) momentum, and the path integral measure $\mathcal {D}\pi$ is defined as well as for $\phi$.
The point is, if we define $H_\text{HMC}[\phi,\pi] = S[\phi] + \frac{1}{2} \pi^2$,
we can regard it as a fictitious classical statistical system with a Hamiltonian $H_\text{HMC}$.

In the other words, realization (joint) probability for a certain configuration $\phi$ and $\pi$ is given by, 
\begin{align}
p_\text{HMC}[\phi,\pi]= \frac{e^{-H_\text{HMC}[\phi,\pi]} }{\mathcal{Z}_\text{HMC}  },
\end{align}
where $\mathcal{Z}_\text{HMC} = \int \mathcal{D} \phi \mathcal{D}\pi \; e^{-H_\text{HMC}[\phi,\pi]} $.
We can easily notice the equivalence
between
$p_\text{HMC}[\phi,\pi]$ and $p_\text{lat}[\phi]$ by integrating out this $\pi_n$ field,
which is not included in obsevables.
Namely marginalization from the joint probability $p_\text{HMC}[\phi,\pi] $ to $p_\text{lat}[\phi]$.

HMC consists of three steps:
\begin{enumerate}
\item {\it Momentum refreshment}: Generate a set of $\pi_n$ for every points $n$ from the Gaussian distribution. \label{item:appendix_HMC_HB}
\item {\it The molecular dynamics}: Fields $(\phi, \pi)$ are evolved to $(\phi', \pi')$ using the canonical equations of motion for $H_\text{HMC}[\phi,\pi] = S[\phi] + \frac{1}{2} \pi^2$ with the {\it leapfrog integrator}
\label{item:appendix_HMC_MD}
\item {\it The Metropolis test}: Chooses a next pair from candidates $(\phi', \pi')$ and $(\phi, \pi)$ using $H_\text{HMC}[\phi,\pi]$. \label{item:appendix_HMC_Met}
\end{enumerate}
This algorithm has been proved to converge to exact path integral \cite{DUANE1987216} thorough the detailed balance, which is proved from the reversibility of each step.
By this reason, we need to utilize a reversible integrator, which is explained later, to solve the equation of motion.

The leapfrog integrator is the simplest symplectic integrator with reversibility.
This is a integrator for solving equations of motion and they can well preserve the value of the Hamiltonian of the system.
First we introduce integrators for $\phi_n$ and $\pi_n$ by, 
\begin{align}
\hat{T}_Q(\tau): \phi_n(0)
=\phi_n \to \phi_n' =\phi_n(\tau), \\
\hat{T}_P(\tau): \pi_n(0)=\pi_n \to \pi_n' = \pi_n(\tau),
\end{align}
where $\tau$ is a fictitious time of the evolution. Each integration is done by the Euler's integration.
This time evolution is determined by solving the canonical equations of motion for $H_\text{HMC}$,
\begin{align}
\frac{d \phi_n}{d \tau}=\frac{\partial H_\text{HMC}}{\partial \pi_n},
\;\;\;
\frac{d \pi_n}{d \tau}= -\frac{\partial H_\text{HMC}}{\partial \phi_n}.
\end{align}
The leapfrog integrator is,
\begin{align}
\hat{T}^\text{Leapfrog}_{QPQ}(\tau) = \left( \hat{T}_Q(\Delta\tau/2) \hat{T}_P(\Delta\tau) \hat{T}_Q(\Delta\tau/2) \right)^{\tau / \Delta\tau},
\end{align}
where $\tau / \Delta\tau $ is a positive integer and conventionally $\tau$ is chosen to 1.
By performing the integrator $\hat{T}^\text{Leapfrog}_{QPQ}(\tau)$, we obtain 
a candidate configuration $(\phi,\pi)(\tau=1)$ from $(\phi,\pi)(\tau=0)$.
However, even if we use a symplectic integrator,
the value of $H_\text{HMC}$ cannot be exactly preserved during
the fictitious time evolution.
Namely $H_\text{HMC} $ varies to $ H_\text{HMC} + O(\Delta \tau^2)$ during the evolution
(see for example \cite{Takaishi:2005tz}).
This is due to violation of the low of energy conservation of a numerical solution.
Thus we need to perform the Metropolis test in step \ref{item:appendix_HMC_Met}
to obtain correct distribution of $e^{-H_\text{HMC}[\phi,\pi] }$.
By repeating step 1 to 3, we obtain a sequence of configurations $\phi$
which obeys $e^{-H_\text{HMC}[\phi,\pi]} \sim e^{-S[\phi]}$.

As we have mentioned, a sequence of configurations are
affected by the autocorrelation, which is evaluated by the autocorrelation function.
The approximated autocorrelation function
\cite{Wolff:2003sm, Madras:1988ei} is defined by,
\begin{align}
\bar\Gamma(\tau) = \frac{1}{\nconf - \tau} \sum_{c}^\nconf (O_c-\bar{O})(O_{c+\tau}-\bar{O}),
\label{autocorrelation_function}
\end{align}
where $O_c = O[ \phi^{(c)} ]$ is the value of operator $O$ for $c$-th configuration $\phi^{(c)}$ and $\tau$ is the fictitious time of HMC.
$\nconf$ is the number of configurations.
We use normalized one, $\bar\rho(\tau)=\bar\Gamma(\tau) /\bar\Gamma(0) $.

The integrated autocorrelation time $\tau_\text{int}$ quantifies 
effects of autocorrelation, which is given by,
\begin{align}
\tau_\text{int} = \frac{1}{2} + \sum^{W}_{\tau=1} \bar\rho(\tau).
\label{tau_int}
\end{align}
Here s window $W$ is set to the first time for large $\tau$ where
\begin{align}
\bar\rho(\tau) \leq \av{ \delta \rho(\tau)^2 }^{1/2},
\end{align}
as in \cite{Luscher:2005rx}.
The error of integrated autocorrelation time
is estimated by the Madras--Sokal formula \cite{Madras:1988ei, Luscher:2005rx},
\begin{align}
\av{\delta \tau_\text{int}^2 }
\simeq
\frac{4W + 2}{\nconf} \tau_\text{int}^2.
\end{align}

\section{Appendix C. GRBM \label{app:rbm}}
In this appendix, we explain Gaussian-Bernoulli Restricted Boltzmann Machines (GRBM) and their update.
In contrast to binary Boltzmann machines, GRBM can treat real-valued inputs.
It consists of
{\it visible} continuous variables, $\phi = \{ \phi_n \}$, and {\it hidden} binary variables, $h =\{h_i\}$.
Basically, GRBM is defined as a statistical physics system controlled by joint probability, 
\begin{align}
p_\theta[\phi, h] =
\frac{e^{- H_\theta^\text{GRBM} [\phi, h] }}{ \mathcal{Z}_\theta},
\end{align}
where 
$\mathcal{Z}_\theta = \int \mathcal{D}\phi\sum_{h} e^{- H_\theta^\text{GRBM} [\phi, h] }$
with the ``Hamiltonian'' for this system \eqref{RBM_H},
\begin{align*}
H^\text{GRBM}_\theta[\phi, h]
=
\sum_{n}
\frac{(\phi_n - a_n)^2}{2 \sigma_n^2}
-
\sum_{j}
b_j h_j
-
\sum_{n, j}
\frac{\phi_n}{\sigma_n} w_{nj} h_j,
\end{align*}
where $\theta = (a_n, \sigma_n, b_j, w_{nj})$ represents parameters to be updated in the learning/training process.
Once we fix parameters $\theta$, one can easily calculate conditioned probabilities as
\begin{align}
&p ( \phi | h) = \prod_n \mathcal{N} \Big( a_n + \sigma_n \sum_j w_{nj} h_j \Big| \sigma_n^2 \Big),
\label{htov}
\\
&
\left. \begin{array}{ll}
p(h_j=1 | \phi) = \sigma \Big( \sum_n \frac{\phi_n}{\sigma_n} w_{nj} + b_j \Big), & \\
p(h_j=0 | \phi) = 1-\sigma \Big( \sum_n \frac{\phi_n}{\sigma_n} w_{nj} + b_j \Big), & \\
\end{array} \right.
\label{vtoh}
\end{align}
and $p(h|\phi) = \prod_j p(h_j | \phi)$.
$\mathcal{N} (\mu | \sigma^2)$ is Gaussian distribution and $\sigma(x) = {1}/{(1+e^{-x})}$.
By using these conditioned probabilities, one can define the following sampling procedure called block Gibbs sampling:
\begin{align}
\phi^\text{GRBM}
\overset{p(\phi_\text{GRBM} | h)}{\longleftarrow}
h
\overset{p(h|\phi)}{\longleftarrow}
\phi.
\label{BG}
\end{align}
For a given sampled data $\phi$, we update $\theta$ by
\begin{align}
\left. \begin{array}{ll}
a_n \leftarrow a_n +  \frac{\epsilon}{\sigma_n^2} (\phi_n - \phi^{\text{GRBM}}_{n}),  \\
b_j \leftarrow b_j + \epsilon \Big[ p(h_j=1 | \phi) - p(h_j=1|\phi^\text{GRBM}) \Big],
\\
w_{nj} \leftarrow w_{nj} 
\\ \qquad \quad
+ \frac{\epsilon}{\sigma_i}  \Big[ \phi_n
p(h_j=1 | \phi) -  \phi_n^\text{GRBM}p(h_j=1 | \phi^\text{GRBM}) \Big], 
\\
\sigma_n \leftarrow (1- \eta) \sigma_n 
\\ \qquad \quad
- \frac{\epsilon \phi_n}{4} \sum_j w_{nj} \Big[
p(h_j=1 | \phi) - p(h_j=1|\phi^\text{GRBM}) \Big]. \\ 
\end{array} \right.,
\label{learning}
\end{align}
where $\epsilon$ and $\eta$ are small values.
It is known that, after iterating the above parameter updates using configurations from HMC
$\{ \phi \}$, the sampling \eqref{BG} with updated parameter provides approximate sampler for actual distribution ${e^{-S [\phi] }}/{\mathcal{Z}_\text{lat}}$ \cite{hinton2006training, bengio2009justifying, cho2011improved}.


\end{document}